\documentclass[%
 reprint,
 amsmath,amssymb,
 aps,
 prl,
]{revtex4-2}

\usepackage{graphicx}
\usepackage{dcolumn}
\usepackage{bm}
\usepackage{color}

\usepackage{cleveref}
\usepackage{afterpage}
\usepackage{float}
\usepackage{amsmath}
\usepackage{bm}

\begin{document}

\preprint{APS/123-QED}

\title{Inertia-driven propulsion of asymmetric spinner-dimers at moderate Reynolds numbers}

\author{Zaiyi Shen}
\affiliation{
State Key Laboratory for Turbulence and Complex Systems, Department of Mechanics and Engineering Science, College of Engineering, Peking University, Beijing 100871, China; E-mail: zaiyi.shen@pku.edn.cn
}

\author{Dongfang Fu}
\affiliation{
State Key Laboratory for Turbulence and Complex Systems, Department of Mechanics and Engineering Science, College of Engineering, Peking University, Beijing 100871, China
}
\author{Juho Lintuvuori}%
\affiliation{%
 Univ. Bordeaux, CNRS, LOMA, UMR 5798, F-33400 Talence, France 
}%

\date{\today}

\begin{abstract}

We investigate the translational motion of rotating colloidal systems at moderate Reynolds numbers (\(\mathrm{Re}\)), focusing on {\color{black} particle dimers in} snowman-like configurations under three  scenarios: (i) two co-rotating spheres driven by an external field, (ii) two counter-rotating spheres driven by an internal torque as a swimmer, and (iii) a single rotating spinner {\color{black} with a} passive sphere for cargo delivery, using hydrodynamic simulations. {\color{black} In all the three cases, the particles are bound together hydrodynamically, and the purely rotational motion of the spinners produces a net propulsion of the dimers along the axis of rotation due to a symmetry breaking.} 
{\color{black} We demonstrate tunable dynamics, where the propulsion direction of the co-rotating dimer can be reversed by tuning the aspect ratio and Reynolds number, as well as cargo transport where a dimer consisting of a single spinner and a passive cargo particle can have a sustained locomotion due to broken head-to-tail symmetry of the overall flow fields.}
These findings {\color{black} highlight} the critical role of inertia in {\color{black} creating locomotion from rotational motion} and offer new avenues for controlling and optimizing translational motion in colloidal assemblies through rotational degrees of freedom.

\end{abstract}

\maketitle

\section{Introduction}
Biological organisms such as bacteria, algae, and spermatozoa employ specialized propulsion strategies to navigate fluid environments~\cite{lauga2009hydrodynamics}. At the micro-scale, where the Reynolds number (\(\mathrm{Re}\)) is typically very low, these swimmers rely on non-reciprocal motions~\cite{lauga2016bacterial,dreyfus2005microscopic}, such as flagellar beating or ciliary strokes, to achieve net movement under conditions dominated by viscosity. However, as the physical size of a swimmer or its speed increases, inertial effects become significant, leading to fundamentally different propulsion mechanisms under higher \(\mathrm{Re}\) conditions~\cite{klotsa2015propulsion,derr2022reciprocal}.

In recent years, a growing interest in artificial micro-swimmers has highlighted the importance of understanding and harnessing fluid inertial effects~\cite{klotsa2019above,lowen2020inertial,chen2024self}. Many synthetic designs aim to replicate or abstract biological propulsion methods~\cite{klumpp2019swimming}. A particularly promising strategy, in both biological and artificial systems, is to exploit rotational degrees of freedom for propulsion~\cite{purcell1997efficiency,rodenborn2013propulsion,bricard2013emergence,kaiser2017flocking,driscoll2017unstable,alapan2020multifunctional}. Rotating structures, such as bacterial flagella or magnetically driven colloids, can generate pronounced flow fields, especially when operating in regimes not strictly confined to \(\mathrm{Re} \ll 1\)~\cite{grzybowski2000dynamic,goto2015purely,fang2020magnetic,shen2023collective}. Indeed, as the {\color{black} particle size} increases or rotational frequencies become large, secondary flows arise and enable net motion in relatively simple, axisymmetric geometries, that would remain stationary under purely viscous conditions (\(\mathrm{Re} \ll 1\))~\cite{nadal2014rotational,chen2024self}. Specifically, breaking head-to-tail symmetry along the spinning axis, has been shown to lead to self-propulsion for a co-rotating colloidal snowman dimer~\cite{nadal2014rotational} and slightly asymmetric cylinders~\cite{chen2024self}, when inertial flows are included. Deepening our understanding of these fluid–structure interactions offers new avenues for micro- and meso-scale robotics, targeted cargo transport, and biomedical applications.

In this paper, we explore how inertial effects enable propulsion in rotating colloids, bridging the gap between the classical Stokes-limit scenario and the more complex flow regimes at moderate Reynolds numbers (\(\mathrm{Re}\)). We begin by {\color{black} validating our simulation method} by examining a single rotating sphere. {\color{black} The result highlights} the transition from purely azimuthal flows at low \(\mathrm{Re}\) to a regime where secondary flows become significant {\color{black} and agree well with theoretical calculations~\cite{bickley1938lxv}}. {\color{black} Although a single spinning sphere remains stationary} across all \(\mathrm{Re}\) due to its inherent symmetry, {\color{black} it has been demonstrated theoretically and numerically~\cite{nadal2014rotational}} that including an additional sphere of a different size, can induce net motion along the spinning axis and towards the larger sphere for $\mathrm{Re}$ up to $12$. 

{\color{black} We build on this, and consider three distinct configurations: (i) co-rotating spheres at {\color{black} $\mathrm{Re}\approx 0\ldots 65$}, (ii) a snowman configuration of counter-rotating spinners and (iii) a driven sphere and a passive cargo particle. In all of the three cases, a spontaneous motility is observed to arise from inertial hydrodynamic flows. Specifically we show that when the Reynolds number is increased, the co-rotating dimer can reverse its direction of motion, and move towards the smaller sphere. Generally,} our findings demonstrate how {\color{black} the size asymmetry between the spheres and inertial effects alter the flow field and drive self-propulsion.}

By systematically varying the Reynolds number (\(\mathrm{Re}\)) and geometric parameters, we identify two primary propulsion regimes: at lower \(\mathrm{Re}\), {\color{black} fluid is drawn in from the spinner poles and pulls the dimer forward. Conversely, at higher \(\mathrm{Re}\), increasingly strong jets emerging from the equatorial region push the spinner ahead. As a summary,} this study provides a comprehensive study of how rotating colloidal systems transition to inertia-dominated propulsion. It offers valuable insights for the design and optimization of future generations of active particle systems capable of operating across diverse fluidic regimes.

\begin{figure*}
\centering
  \includegraphics[width=2\columnwidth]{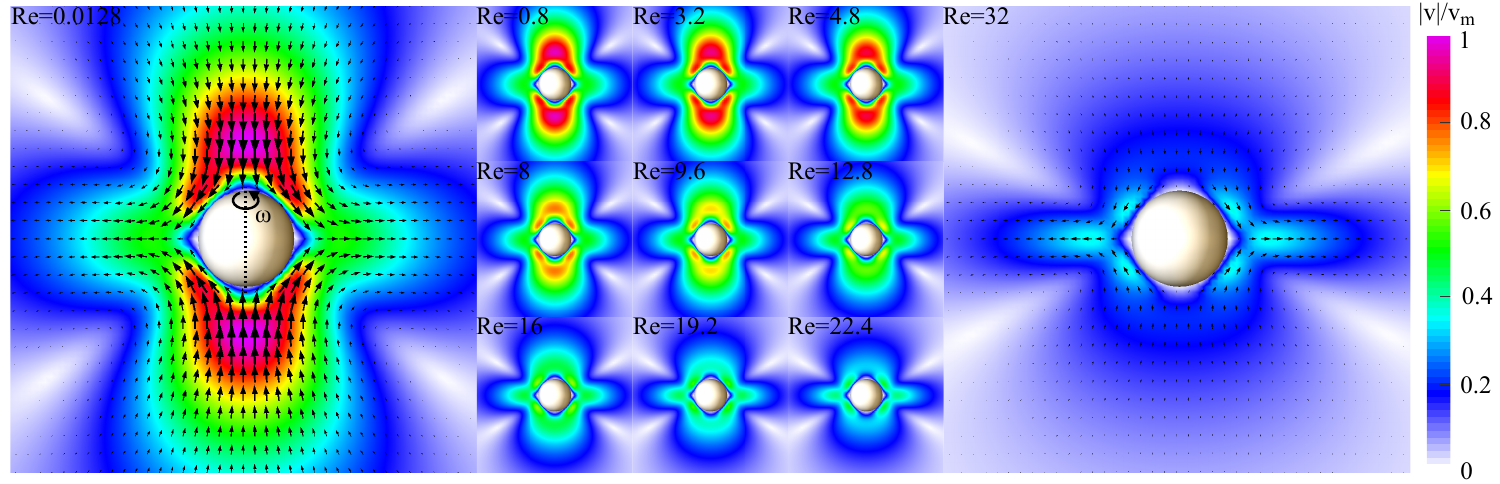}
  \caption{
  The flow field in the meridional plane for a spherical particle rotating at Reynolds numbers (\(\mathrm{Re}\)) ranging from 0.01 to 32. The color map represents the magnitude of the secondary flow normalized by the maximum radial velocity obtained from the asymptotic solution.
  }
  \label{second}
\end{figure*}

\begin{figure*}
 \centering
 \includegraphics[width=2\columnwidth]{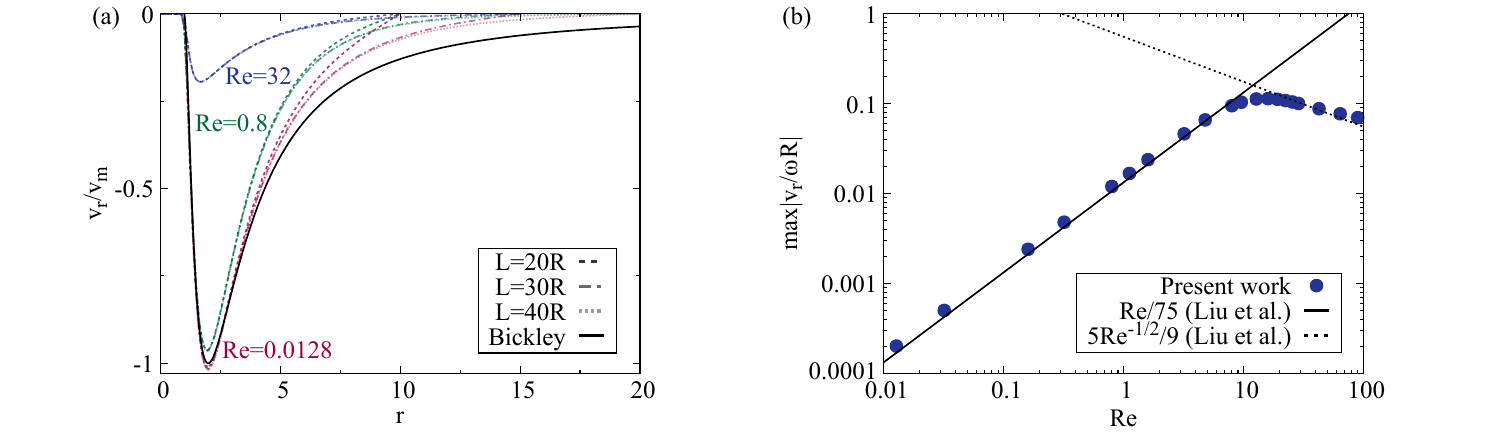}
\caption{
(a) The normalized radial component of the fluid velocity along the spinning pole for \(\mathrm{Re} = 0.0128, 0.8, 32\). For each \(\mathrm{Re}\), three curves are plotted, corresponding to simulations with computational domain sizes of \(L = 20R\), \(30R\), and \(40R\). The black solid line represents the asymptotic solution.
(b) The maximum radial velocity along the spinning pole is plotted as a function of \(\mathrm{Re}\). The lines correspond to results from previous numerical simulations~\cite{liu2010wall}.
}
\label{pf}
\end{figure*}

\section{Numerical methods and validation}
We employ the open-source package "Ludwig"~\cite{kevinstratford_2024_12822477} to implement the lattice Boltzmann method (LBM) for simulating the system dynamics. Fluid–particle interactions are handled using the bounce-back on links method, ensuring a no-slip boundary condition on the particle surfaces. To prevent particle–particle overlap, a short-range repulsive force is applied between solid boundaries, following the approach outlined in relevant references~\cite{shen2018hydrodynamic,shen2019hydrodynamic}. The cutoff of the repulsion is {\color{black} $0.5\Delta x$, where $\Delta x$ is the lattice spacing.}

All simulations for the spinner-dimer are conducted within a cubic simulation box of side length \(L = 20R\). In the chosen simulation units, {\color{black} where the larger spinner has a fixed radius of $R = 8\Delta x$}, while the radius of the second spinner is varied to introduce geometric asymmetry. The fluid density is set to \(\rho = 1\).

We investigate the system at finite rotational Reynolds numbers, defined as \(\mathrm{Re} = \rho R^2 \omega/ \mu\), which quantify the ratio of inertial to viscous forces. Here, \(\omega\) represents the angular velocity of the spinner, and \(\mu\) denotes the fluid viscosity, which is adjusted to achieve the desired \(\mathrm{Re}\) values.

In the Stokes limit \((\mathrm{Re} \ll 1)\), a rotating sphere generates only azimuthal flow. However, as the Reynolds number increases, a secondary flow begins to emerge~\cite{climent2007dynamic,shen2020hydrodynamic}. Due to inertial effects, the flow converges toward the poles and moves outward near the equator. An asymptotic analysis for small \(\mathrm{Re}\) \((\mathrm{Re} \ll 1)\) has been derived~\cite{bickley1938lxv,climent2007dynamic}, yielding:
\begin{equation}
  v_\theta(r)=\frac{\omega R^3}{r^2}\sin{\psi}+O(\mathrm{Re^2}),
\end{equation}
\begin{equation}
  v_r(r)=-\frac{\omega R^3}{8 r^2}(3\cos^2{\psi}-1)(1-\frac{R}{r})^2 \mathrm{Re}+O(\mathrm{Re^2}),
\end{equation}
\begin{equation}
  v_\psi(r)=\frac{\omega R^4}{4 r^3}(1-\frac{R}{r})\sin{\psi}\cos{\psi} \mathrm{Re}+O(\mathrm{Re^2}).
\end{equation}

These equations describe the azimuthal, radial, and polar velocity components of the flow induced by the rotating sphere, respectively. To validate our numerical results, we compare them against these analytical solutions. Fig.~\ref{second} shows the velocity field in the meridional plane for various Reynolds numbers. At low \(\mathrm{Re}\) (e.g., \(\mathrm{Re} = 0.0128\)), the flow field closely matches the analytical solution for the secondary flow (Fig.~\ref{second}), and this agreement remains good for \(\mathrm{Re} < 5\), with only minor deviations.

As \(\mathrm{Re}\) increases, the asymptotic assumptions become invalid, leading to significant changes in the flow behavior. A marked difference from the cross-sectional slice of the velocity field (the full flow is axisymmetric) is that the incoming flow from the polar region tends to split into two distinct streams, while the outward flow intensifies at the equator, forming a jet-like structure (Fig.~\ref{second}). This effect can be attributed to the growing influence of inertia (like centrifugal effects) with increasing \(\mathrm{Re}\).  

To quantitatively validate our numerical results, we calculated the velocity profile along the spinning pole. For \(\mathrm{Re} = 0.0128\) and \(\mathrm{Re} = 0.8\), the near-field velocity exhibits good agreement with the asymptotic solution (Fig.~\ref{pf} a). However, at \(\mathrm{Re} = 32\), the flow profile deviates from the analytical prediction because the Reynolds number is too high for the asymptotic assumptions to remain valid. In the far field, the numerical results show a faster decay compared to the analytical solution (Fig.~\ref{pf} a), which can be attributed to the periodic boundary conditions used in the simulations. 
Additionally, we determined the maximum velocity along the rotation axis and compared our data with previous studies that used other numerical methods~\cite{liu2010wall}. The resulting velocity-versus-\(\mathrm{Re}\) curve shows excellent agreement with these earlier findings (Fig.~\ref{pf} b).

For a single spherical particle, the system maintains axisymmetry about the rotational axis and mirror symmetry across the equatorial plane. As a result, no translational motion of the sphere is expected, even when the flow becomes nonlinear at higher Reynolds numbers. However, when two spheres of different sizes form a snowman-like configuration with co-rotating axes, the dynamics change significantly {\color{black} and the dimer can self-propel along its spinning axis towards the larger sphere~\cite{nadal2014rotational}.}

At small Reynolds numbers, the flow remains predominantly azimuthal, preserving the system's symmetry and preventing any translational motion. As the rotational frequency (and thus the Reynolds number) increases, secondary flows emerge. The size difference between the two spheres breaks the equatorial-plane symmetry, triggering the onset of translational motion of the spinner-dimer. We systematically explore this behavior in three representative scenarios: (i) two co-rotating spheres driven by an external field, (ii) two counter-rotating spheres driven by an internal torque as a swimmer, and (iii) a single rotating spinner attached to a passive sphere for cargo transport.

\section{Results}

\subsection{Two co-rotating spheres}

\begin{figure*}
\centering
\includegraphics[width=2\columnwidth]{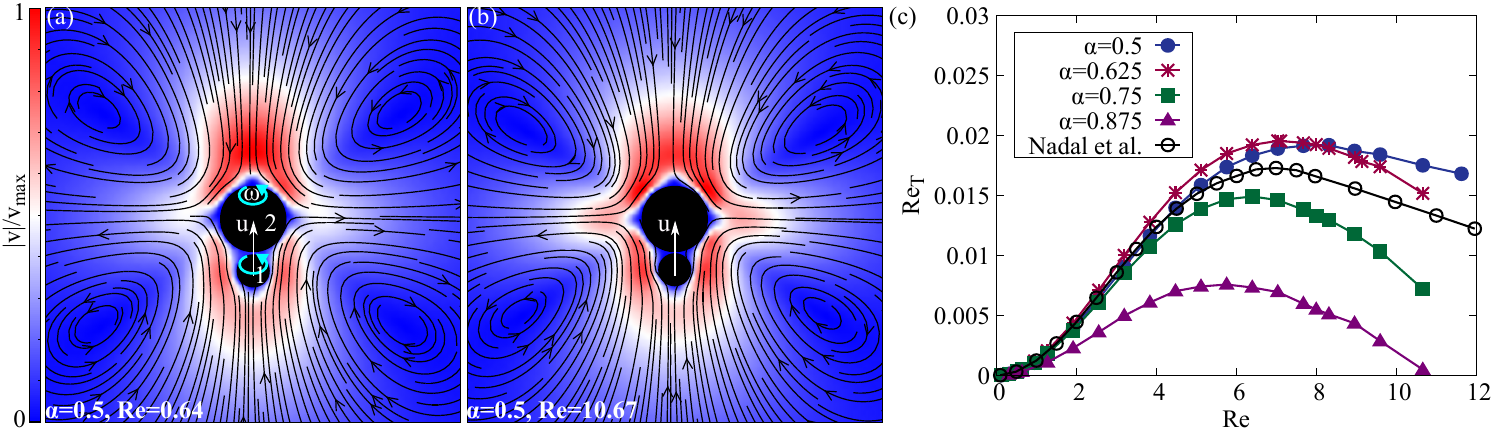}
\caption{
Two coaxial spherical particles rotating at the same angular velocity \(\omega\) can achieve translational motion along their common axis. The streamlines for (a) \(\mathrm{Re} = 0.64\) and (b) \(\mathrm{Re} = 10.67\) are shown for a size ratio of \(\alpha = 0.5\). 
(c) The translational Reynolds number \(\mathrm{Re}_T\) is plotted as a function of the rotational Reynolds number \(\mathrm{Re}\) for various values of \(\alpha\). The hollow circles represent the case of \(\alpha = 0.5\), as obtained by Nadal et al.~\cite{nadal2014rotational} using the finite element method in a moving reference frame. All curves exhibit the presence of an optimal Reynolds number, \(\mathrm{Re} \approx 7\), which corresponds to the maximum \(\mathrm{Re}_T\).
}
\label{fl}
\end{figure*}

\begin{figure*}
\centering
\includegraphics[width=2\columnwidth]{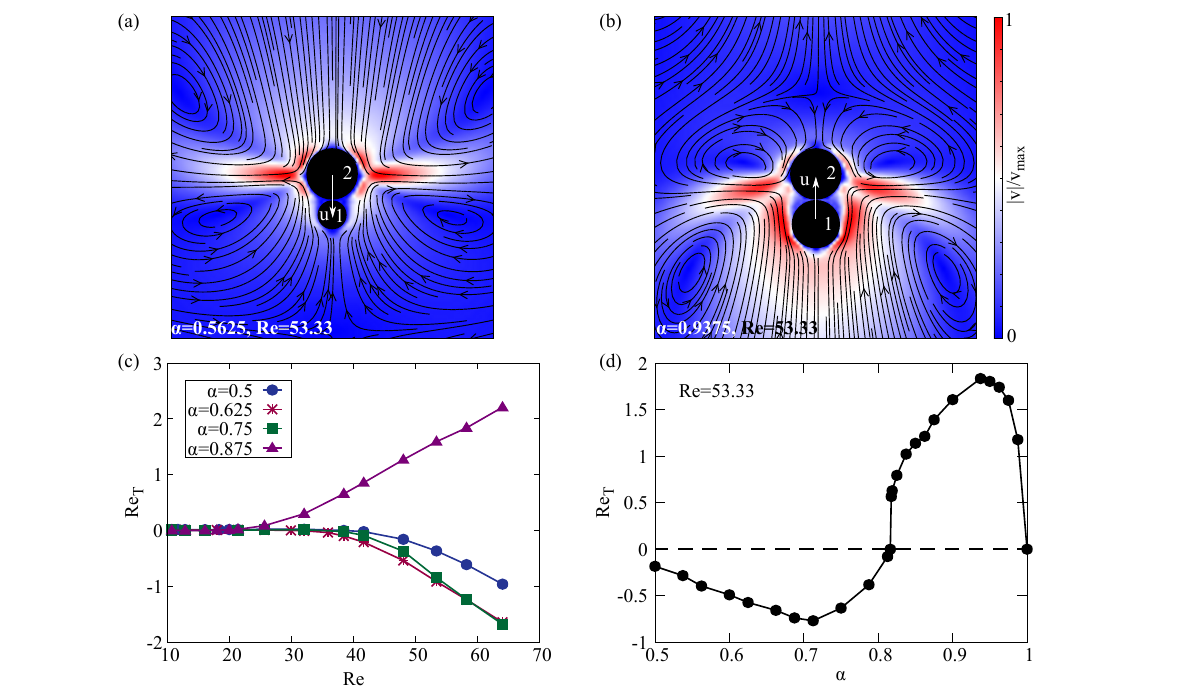}
\caption{
(a) The dimer moves downward (from the larger sphere to the smaller one) when \(\alpha = 0.5625\) at \(\mathrm{Re} = 53.33\).  
(b) The dimer moves upward (from the smaller sphere to the larger one) when \(\alpha = 0.9375\) at \(\mathrm{Re} = 53.33\).  
(c) The translational Reynolds number \(\mathrm{Re}_T\) is plotted as a function of the rotational Reynolds number \(\mathrm{Re}\) for various values of \(\alpha\).  
(d) The translational Reynolds number \(\mathrm{Re}_T\) is plotted as a function of \(\alpha\) at a fixed rotational Reynolds number of \(\mathrm{Re} = 53.33\). 
}
\label{fh}
\end{figure*}

First, we examine the translational motion of two spheres with co-rotating axes and identical rotational frequencies, forming a snowman-like colloid driven by a rotating field along their common axis. {\color{black} The configuration is similar to ref.~\cite{nadal2014rotational}, but we extend the exploration to larger Reynolds numbers and systematically vary the aspect ratio between the sizes of the two spheres.} We define the rotational Reynolds number (\(\mathrm{Re}=\rho R^2 \omega / \mu\)) and translational Reynolds number (\(\mathrm{Re_T}=\rho R u / \mu\)) using the radius of the larger particle (\(R = R_2\)) in the dimer (Fig.~\ref{fl} a). The aspect ratio is given by \(\alpha = R_1 / R_2\), where \(R_1\) and \(R_2\) are the radii of the smaller and larger spheres, respectively.

At relatively small Reynolds numbers, the secondary flow closely follows the asymptotic behavior. Applying the analytical solution of the secondary flow field, we see that the flow velocity near the poles increases with \(\mathrm{Re}\) when the angular velocity ($\omega$) is fixed. Consequently, the incoming flow at the front of the larger sphere where \(\mathrm{Re}\) is higher becomes more intense, creating a stronger pulling on the dimer. This imbalance induces translational motion from the smaller sphere toward the larger one, as illustrated in Fig.~\ref{fl}~a, {\color{black} and in agreement with previous studies~\cite{nadal2014rotational}.}

\begin{figure*}
\centering
\includegraphics[width=2\columnwidth]{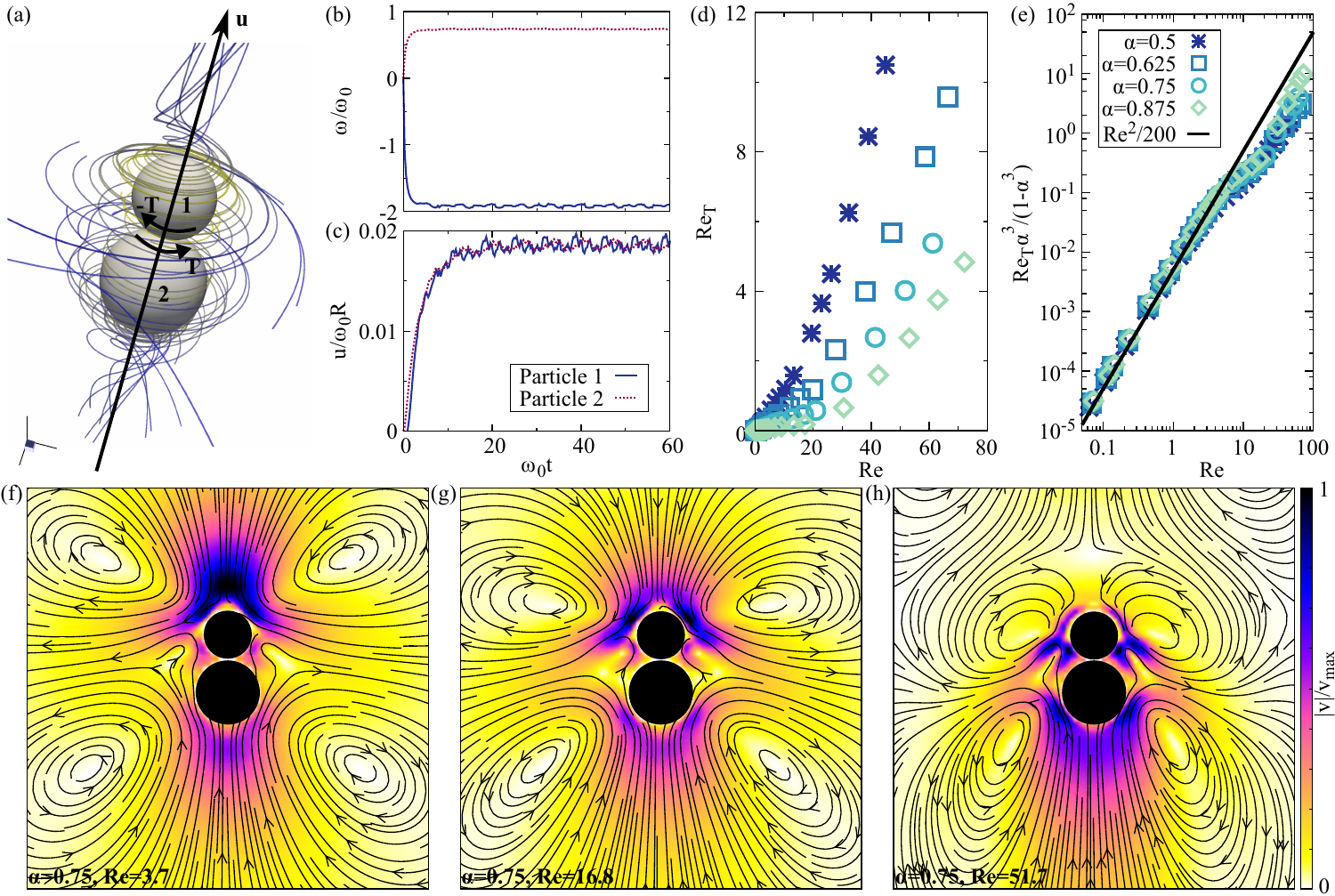}
\caption{
(a) A schematic of a swimmer composed of two coaxial spinners driven by equal and opposite internal torques. When the two spinners differ in size and rotate at finite Reynolds numbers, the swimmer translates along the direction indicated by the arrow. The streamlines illustrate a typical flow field around the swimmer.
(b) The angular velocity and (c) the translational velocity of each particle in the swimmer, for an aspect ratio \(\alpha = 0.75\) and \(\mathrm{Re} \approx 5\).
(d) The translational Reynolds number \(\mathrm{Re}_T\) as a function of the rotational Reynolds number \(\mathrm{Re}\) for various aspect ratios \(\alpha\).
(e) The same \(\mathrm{Re}_T\) data, rescaled by the factor \(\alpha^3 / (1 - \alpha^3)\), plotted against \(\mathrm{Re}\).
(f) The flow field generated by the swimmer at different \(\mathrm{Re}\) values for \(\alpha = 0.75\).}
\label{sf}
\end{figure*}

This qualitative understanding holds for relatively small Re (Re<10) even the secondary flow for a single particle starts to deviate from the analytical solution when Re is larger than 5 {\color{black} (see {\it e.g.} Fig.~\ref{second})}. Indeed, the flow fields generated by the dual spinners at \(\mathrm{Re} = 0.64\) and \(\mathrm{Re} = 10.67\) exhibit similar patterns (Fig.~\ref{fl} a and b). For an aspect ratio of \(\alpha = 0.5\), our results show good agreement with previous studies~\cite{nadal2014rotational} ({\color{black} closed blue and open black circles in} Fig.~\ref{fl} c). {\color{black} The translational motion is evident from the increase of the} translational Reynolds number \(\mathrm{Re_T}\), which increases with \(\mathrm{Re}\) (Fig.~\ref{fl} c). {\color{black} It reaches} a maximum value around \(\mathrm{Re} = 7\), after which it begins to decline . At higher Reynolds numbers, our simulations show deviations from~\cite{nadal2014rotational}, which may {\color{black} likely to} be attributed to factors such as periodic boundary effects, numerical resolution, or the surface distance between the spheres.  

{\color{black} To gain insight to the effects of the dimer geometry,} we also conducted simulations for additional aspect ratios {\color{black} $\alpha\approx 0.5\ldots 0.875$ (Fig.~\ref{fl}c)}. In all the cases, {\color{black} we observed directional movement along the spinning axis and in the direction given by the larger sphere. Furthermore, it was observed that the optimal spinning frequency for locomotion corresponds to Reynolds number of approximately \(\mathrm{Re} = 7\) for all aspect ratios considered (Fig.~\ref{fl} c).} Furthermore, our results show that when the aspect ratio exceeds 0.625, the translational velocity decreases (Fig.~\ref{fl} c), suggesting that the optimal \(\alpha\) for maximum translational performance lies between 0.5 and 0.625. This behavior can be understood through symmetry considerations. {\color{black}  A small  $\alpha$} makes the configuration more similar to a single sphere {\color{black} with a radius $R_2$}, while a larger \(\alpha\) gives the system a dumbbell-like shape {\color{black} $R_1\approx R_2$}. {\color{black} At both limits, increasing the dimer} symmetry reduces the flow asymmetry between the front and rear of {\color{black} of the snowman dimer. This leads to reduction} of its net translational velocity.

{\color{black} Interestingly, the data in Fig.~\ref{fl}c {\color{black} show} that when \(\mathrm{Re}\) increases well beyond $\mathrm{Re}\sim 7$ the translational velocity decreases, eventually reaching zero for $\alpha\approx 0.875$ around $\mathrm{Re}\approx 11$. This suggests the possibility {\color{black} of reversing the direction of locomotion.}} 

{\color{black} To {\color{black} further investigate} the effects of the aspect ratio $\alpha$ and the spinning frequency,}, we extend our simulations up to {\color{black} \(\mathrm{Re} \approx 65\)} (Fig.~\ref{fh}). {\color{black} The simulations reveal that} the direction of motion depends on the aspect ratio \(\alpha\) {\color{black} for a given Reynolds number}. For \(\mathrm{Re} > 20\), the {\color{black} spinner moves in the direction from the smaller sphere towards the larger one when \(\alpha \approx  0.875\), {\color{black} while, for smaller values of \(\alpha\)}, the motion reverses its direction (Fig.~\ref{fh}c). }

Unlike in lower \(\mathrm{Re}\), where {\color{black} the single particle} secondary flows dominate, the flow at higher \(\mathrm{Re}\) becomes more complex and is {\color{black} primarily driven} by a jet originating near the equator of the larger spinner (Fig.~\ref{fh} a and b). This jet pushes the dimer forward, and its direction depends strongly on the geometry. For a smaller aspect ratio \(\alpha\), {\color{black} the jet is located at the equatorial region of the larger particle and slightly tilted upward, away from the smaller particle (red region in Fig.~\ref{fh}a). For larger \(\alpha\), two eddies appear near the larger spinner, and the fluid jet at the equatorial region tilts slightly downward (towards the smaller sphere).  This reverses the locomotion direction from the direction towards the smaller particle (Fig.~\ref{fh}a) to motion in the direction of the larger colloid (Fig.~\ref{fh}b), when the aspect ratio is modified for a fixed $\mathrm{Re}\approx 53$.}

A more detailed study at $\mathrm{Re} \approx 53$ reveals a critical \(\alpha \approx 0.82\) at which the velocity direction changes (Fig.~\ref{fh} d). Below this critical value (\(\alpha < 0.82\)), the colloid moves from the larger sphere to the smaller one, attaining a maximum velocity around \(\alpha = 0.7\). Above this threshold (\(\alpha > 0.82\)), the colloid reverses direction and moves from the smaller sphere toward the larger one, reaching a maximum velocity near \(\alpha = 0.94\).

\subsection{Two counter-rotating spheres}

{\color{black} In the second scenario, we consider a torque and force-free configuration. It consists of two spheres, driven by equal but opposite internal torques (Fig.~\ref{sf}a). This makes the dual spinner an effective force-free swimmer.} This configuration has been experimentally and theoretically shown to produce self-locomotion in viscoelastic fluids at zero Reynolds number~\cite{kroo2022freely,binagia2021self}, {\color{black} where more intricate coupling between rotation and translation, as well as other swimmers created stresses can be expected~\cite{goychuk2014molecular,li2021microswimming,theeyancheri2022silico}.}

In this configuration, each sphere experiences an equal and opposite torque \(T\), causing them to rotate in opposite directions (Fig.~\ref{sf}b). Under Stokes flow conditions, {\color{black} this results in an angular velocity $\omega_{1,2}=T/8\pi \mu R^3_{1,2}$, where $R_{1,2}$ corresponds to the radii of particle 1 and particle 2, respectively. Thus the smaller particle is expected to spin faster. This holds reasonably well for the particles in the dimer as well. Considering an aspect ratio $\alpha\equiv R_1/R_2=0.75$ dimer, a ratio between the spinning frequencies $|\omega_2/\omega_1|\approx 2.3$ is observed (Fig.~\ref{sf}b), which agrees well with $|\omega_2/\omega_1| \approx 2.4$ expected for isolated spinners.} 

Consequently, the rotational Reynolds number {\color{black} of an isolated spinner scales as \(\mathrm{Re} \sim T / R\). The corresponding inertial secondary flow of an isolated spinner is given by equations 1-3. The radial component $v_r$ at polar regions, $\cos\psi = 1$, scales as \(v_r \sim \mathrm{Re}^2 / R\).} This implies that for a fixed torque \(T\), the smaller sphere experiences a higher \(\mathrm{Re}\) and generates a stronger secondary flow in front of it, {\color{black} pulling the surrounding fluid towards itself (see {\it e.g.} Fig.~\ref{second}a).} As a result, {\color{black} one can expect that the dual-spinner swimmer translates in the direction of the smaller sphere. This agrees with the simulation results (Fig.~\ref{sf}a). At the steady state, both the spheres are bound together via a mutual attraction arising from the secondary flow at the polar regions, and translate in the direction of the smaller sphere approximately at equal velocities (Fig.~\ref{sf}c).}

A scaling argument based on the secondary flow suggests that the net force driving the swimming motion arises from the difference in the radial flow components at the front and rear of the {\color{black} colloidal dimer}. The swimming speed can be approximated as:

\[
u \sim \left(\frac{\mathrm{Re}_1^2}{R_1} - \frac{\mathrm{Re}_2^2}{R_2}\right).
\]

Expressing this in terms of the translational Reynolds number \(\mathrm{Re}_T = \rho R u/\mu\) and the rotational Reynolds number \(\mathrm{Re}  = \rho R^2 \omega /\mu\), {\color{black} using the radius of the larger sphere (\(R = R_2\)) }, yields:

\[
\mathrm{Re}_T \cdot \frac{\alpha^3}{1 - \alpha^3} \sim \mathrm{Re}^2,
\]

where \(\alpha = R_1 / R_2 < 1\) {\color{black} is } the aspect ratio.

Fig.~\ref{sf}d shows the behavior of \(\mathrm{Re}_T\) {\color{black} measured from the simulations} as a function of \(\mathrm{Re}\) for different values of \(\alpha\). The translational velocity increases with \(\mathrm{Re}\), and greater asymmetry (smaller \(\alpha\)) yields higher swimming speeds (Fig.~\ref{sf}d). When we rescale the plot by \(\alpha^3/(1-\alpha^3)\) and use a log–log format, the data for \(\mathrm{Re} < 5\) collapse onto a single curve (Fig.~\ref{sf}e), supporting our scaling analysis.

At larger \(\mathrm{Re}\), the flow field deviates from the asymptotic solution. As shown in Fig.~\ref{sf}e, the main features related to aspect ratio still hold, but the data indicate two distinct regimes for higher \(\mathrm{Re}\). For \(5 < \mathrm{Re} < 20\), the dimer is pulled by the fluid at the front, though this pulling flow splits into two streams and shifts laterally as \(\mathrm{Re}\) increases (dark regions in Fig.~\ref{sf}g). Once \(\mathrm{Re}\) exceeds about 20, the propulsion mechanism changes. As illustrated in Fig.~\ref{sf}h for \(\mathrm{Re} = 51.7\), jets form on the side of the smaller sphere, pushing the swimmer forward. {\color{black} Contrary to what is observed with the forced snowman dimer (Fig.~\ref{fl}), no reversal of the swimming the direction is observed. The internally driven force-free swimmer continues to swim along the direction given by the smaller sphere.}

\subsection{One rotating and one passive sphere}

\begin{figure*}
\centering
\includegraphics[width=2\columnwidth]{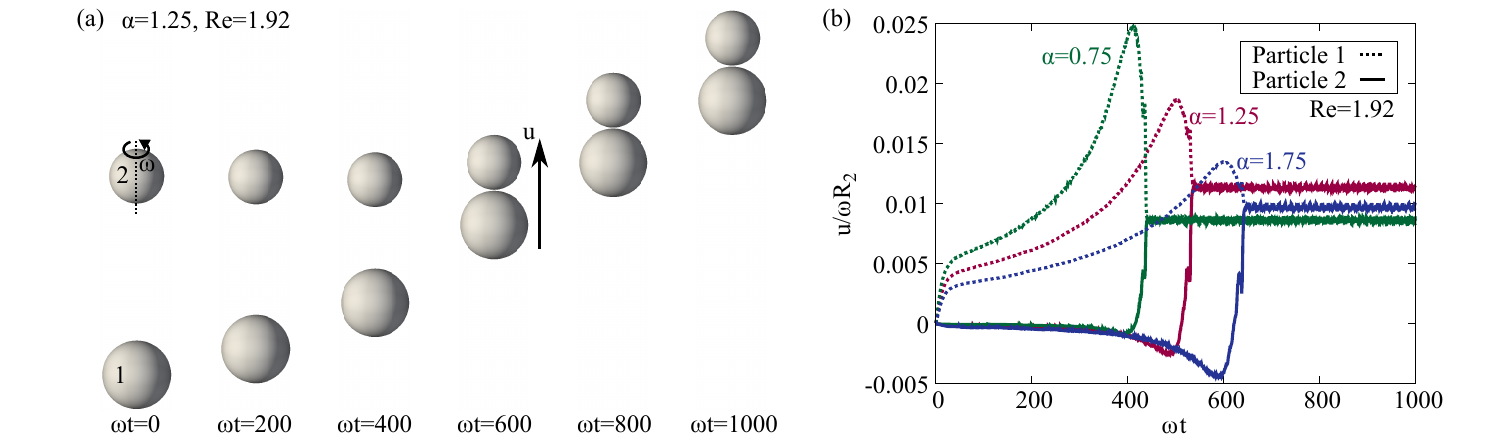}
\caption{
(a) A time series illustrating the positions of two particles when particle 2 is rotating. The passive particle (particle 1) is drawn toward the rotating particle by hydrodynamic interactions. Once they come into contact, the two particles move forward together.
(b) The velocity of the resulting two-particle assembly is shown for different size ratios \(\alpha = 0.75\), \(\alpha = 1.25\), and \(\alpha = 1.75\). In each case, a constant velocity is eventually reached in the final state.
}
\label{p1}
\end{figure*}

{\color{black} Finally we investigate the possibility of cargo transport using the inertial flow fields. We consider a single spinner at finite \(\mathrm{Re}\).} Although a single spinner does not move on its own, {\color{black} including another object can break the symmetry and lead to locomotion. To test this, we placed a passive colloidal particle behind the spinner and along the spinning axis (left panel in Fig.~\ref{p1}a). The spinning particle creates a secondary flow which advects the fluid towards itself at the polar regions. This creates an attraction between the spinner and the cargo. Now the two particles approach each other ($\omega t < 600 $ in Fig.~\ref{p1}a and b).} 
Once a contact is established, the spinner and the passive sphere translate together, moving in the direction from the passive sphere towards the {\color{black}  spinning particle ($\omega t > 600$ in Fig.~\ref{p1}a and b).} This behavior is observed to hold for various sizes of the  passive loads. However, the translational speed of the {\color{black} dimer does not vary} monotonically with the aspect ratio $\alpha$ (the ratio of the passive sphere radius to the spinner radius). {\color{black} For a fixed $\mathrm{Re}\approx 1.9$, the $\alpha\approx 1.25$ dimer is observed to translate at higher speed compared to the $\alpha\approx 0.75$ and $\alpha\approx 1.75$ dimers (Fig.~\ref{p1}b). This suggests that for a given Reynolds number, there exists an optimal payload size.}

\begin{figure*}
\centering\includegraphics[width=2\columnwidth]{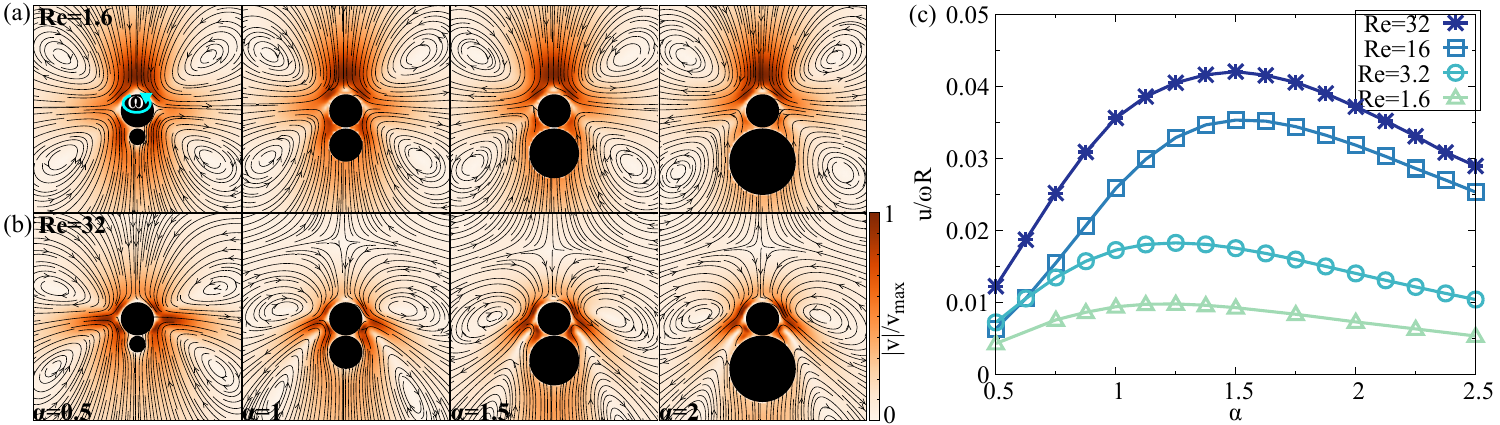}
\caption{
(a) Flow fields at relatively small Reynolds number (\(\mathrm{Re} \approx 1.6\)) for various aspect ratios $\alpha\approx 0.5,~1.0,~1.5,~2.0$.  
(b) Flow fields at relatively large \(\mathrm{Re}\approx 32\) for the same set of aspect ratios.  
(c) The translational velocity in the final steady state, plotted as a function of \(\alpha\) for different values of \(\mathrm{Re}\).
}
\label{p2}
\end{figure*}

{\color{black} To study the effects of the $\alpha$ and Reynolds number for the cargo transport in more detail, we carried out simulations of the hydrodynamically bound dimer for $\alpha \approx 0.5\ldots 2.5$ and $\mathrm{Re}\approx 1.6,~3.2,~16,~32$ (Fig.~\ref{p2}). Increasing the Re was observed to increase the translational speed of the dimer (Fig.~\ref{p2}b). The $\alpha$ dependence shows non-monotonic behaviour with a clear maximum (Fig.~\ref{p2}b). Starting from $\alpha$, initially speed of the dimer increases with increasing $\alpha$, for all the Reynolds numbers considered.  After reaching an optimal aspect ratio $\alpha^*$ the speed is observed to decrease.} Our results show that the maximum velocity appears when \(\alpha\) lies between 1.2 and 1.5, and this optimal value is observed to increase with \(\mathrm{Re}\) (Fig.~\ref{p2}c). {\color{black} One should also note, that locomotion is observed for a $\alpha\approx 1$ dimer, in contrast to the externally and internally driven snowman dimers, where the symmetrical dumbbell is expected to be stationary for symmetry reasons.}

{\color{black} To understand the non-monotonic behaviour, one can consider} two competing effects. On one hand, a larger passive sphere introduces greater asymmetry, potentially enhancing the propulsion. On the other hand, a larger sphere also adds more fluid drag. These opposing influences lead to an optimal translational velocity at an intermediate aspect ratio. 

{\color{black} Finally, the locomotion mechanism is observed to also depend on the Reynolds number.} As in the previous cases, when \(\mathrm{Re}\) is small, the locomotion is dominated by fluid pulled in at the {\color{black} front of the spinner} (Fig.~\ref{p2}a), whereas at higher \(\mathrm{Re}\), jets formed by the spinner provide a pushing mechanism (Fig.~\ref{p2} b). Despite these different propulsion regimes, the optimal velocity still arises from the trade-off between propulsion and hydrodynamic drag.

\section{Conclusions}
Throughout this work, we have explored how rotating colloidal particles, ranging from single spinner to two-sphere assembly, behave under varying Reynolds numbers (\(\mathrm{Re}\)). {\color{black} Our simulations demonstrate how the secondary flow created by spinning particles at finite Reynolds numbers can hydrodynamically bind the particles together. When a size asymmetry between the particles is introduced, a locomotion along the spinning axis can be realised.} 

We started by examining a single rotating sphere, which, due to its symmetry, experiences no net translational motion. As the rotational Reynolds number \(\mathrm{Re}\) increases, secondary flows arise but do not break the inherent symmetry of a single sphere.

When a second {\color{black} spinner or a passive “cargo” is introduced}, however, asymmetry emerges, and the {\color{black} combination becomes motile.} {\color{black} We studied the locomotion in the three distinct configurations: (i) an externally driven snowman dimer, (ii) internally driven force and torque-free swimmer, consisting of two spinner driven by equal but opposite torques, and (iii) a spinner with a passive load particle. We carried out the exploration for various aspect ratios between the particles and wide range of Reynolds numbers $\mathrm{Re}\approx 0\ldots 100$. Our simulations revealed that externally driven snowman dimer can reverse its locomotion direction as a function of the aspect ratio and Reynolds number. Further, a cargo transport was demonstrated using a single spinner where a passive payload was hydrodynamically attached to the spinner. Here, the locomotion arises from the broken head-to-tail symmetry similarly to the externally and internally driven dimers.} 

{\color{black} Typically in experiments, confining walls are present. A single spinning sphere at finite Re is attracted to a flat no-slip wall along its spinning axis due to the inertial hydrodynamic flows~\cite{liu2010wall}. This is likely true for our dimers as well. Near a confining surface, the dynamics is likely dominated by an intricate balance between wall-spinner and spinner-spinner hydrodynamics. This could lead to the dissolution of the hydrodynamically bound snowman dimers. One possibility could be to use solid particles with broken head-to-tail symmetry~\cite{chen2024self}.}

{\color{black} Finally, our simulations revealed two distinct propulsion mechanisms depending on the rotational Reynolds number $\mathrm{Re}$.} At low \(\mathrm{Re}\), the propulsion mechanism is primarily a “pulling” effect {\color{black} where the fluid is advected primarily towards the swimmer at polar regions.} At higher \(\mathrm{Re}\), jet-like flows form {\color{black} at the equatorial region of the spinner}, creating a “pushing” mechanism. Despite this transition, the same fundamental principle {\color{black} for the propulsion} applies : {\color{black} breaking the head-to-tail symmetry, either by size asymmetry, or by introducing a passive load, leads to a different fluid flow at the front and the rear of the dimer, resulting in a} net fluid forcing and, consequently, {\color{black} sustained} propulsion along the spinning axis of the particles. 

Overall, these findings provide insight into designing self-propelled systems and micro-swimmers driven by rotational motion, highlighting how aspect ratio and inertia collectively shape their hydrodynamic behavior.

\section*{Conflicts of interest}
There are no conflicts of interest to declare.

\section*{Data availability}
The Ludwig code used in this article can be found in the repository at https://github.com/ludwig-cf/ludwig/tree/ludwig-0.22.0

\section*{Acknowledgements}
ZS and JSL acknowledge IdEx (Initiative d'Excellence) Bordeaux for funding, Curta cluster for computational time. JSL  acknowledges support by  the  French  National  Research  Agency (ANR) through  Contract No. ANR-19-CE06-0012-01 and la r\'egion Nouvelle Aquitaine project GASPP. ZS acknowledges support by the Fundamental Research Funds for the Central Universities, Peking University. 




\bibliography{rsc} 

\begin{thebibliography}{32}%
\makeatletter
\providecommand \@ifxundefined [1]{%
 \@ifx{#1\undefined}
}%
\providecommand \@ifnum [1]{%
 \ifnum #1\expandafter \@firstoftwo
 \else \expandafter \@secondoftwo
 \fi
}%
\providecommand \@ifx [1]{%
 \ifx #1\expandafter \@firstoftwo
 \else \expandafter \@secondoftwo
 \fi
}%
\providecommand \natexlab [1]{#1}%
\providecommand \enquote  [1]{``#1''}%
\providecommand \bibnamefont  [1]{#1}%
\providecommand \bibfnamefont [1]{#1}%
\providecommand \citenamefont [1]{#1}%
\providecommand \href@noop [0]{\@secondoftwo}%
\providecommand \href [0]{\begingroup \@sanitize@url \@href}%
\providecommand \@href[1]{\@@startlink{#1}\@@href}%
\providecommand \@@href[1]{\endgroup#1\@@endlink}%
\providecommand \@sanitize@url [0]{\catcode `\\12\catcode `\$12\catcode
  `\&12\catcode `\#12\catcode `\^12\catcode `\_12\catcode `\%12\relax}%
\providecommand \@@startlink[1]{}%
\providecommand \@@endlink[0]{}%
\providecommand \url  [0]{\begingroup\@sanitize@url \@url }%
\providecommand \@url [1]{\endgroup\@href {#1}{\urlprefix }}%
\providecommand \urlprefix  [0]{URL }%
\providecommand \Eprint [0]{\href }%
\providecommand \doibase [0]{https://doi.org/}%
\providecommand \selectlanguage [0]{\@gobble}%
\providecommand \bibinfo  [0]{\@secondoftwo}%
\providecommand \bibfield  [0]{\@secondoftwo}%
\providecommand \translation [1]{[#1]}%
\providecommand \BibitemOpen [0]{}%
\providecommand \bibitemStop [0]{}%
\providecommand \bibitemNoStop [0]{.\EOS\space}%
\providecommand \EOS [0]{\spacefactor3000\relax}%
\providecommand \BibitemShut  [1]{\csname bibitem#1\endcsname}%
\let\auto@bib@innerbib\@empty
\bibitem [{\citenamefont {Lauga}\ and\ \citenamefont
  {Powers}(2009)}]{lauga2009hydrodynamics}%
  \BibitemOpen
  \bibfield  {author} {\bibinfo {author} {\bibfnamefont {E.}~\bibnamefont
  {Lauga}}\ and\ \bibinfo {author} {\bibfnamefont {T.~R.}\ \bibnamefont
  {Powers}},\ }\bibfield  {title} {\bibinfo {title} {The hydrodynamics of
  swimming microorganisms},\ }\href@noop {} {\bibfield  {journal} {\bibinfo
  {journal} {Reports on Progress in Physics}\ }\textbf {\bibinfo {volume}
  {72}},\ \bibinfo {pages} {096601} (\bibinfo {year} {2009})}\BibitemShut
  {NoStop}%
\bibitem [{\citenamefont {Lauga}(2016)}]{lauga2016bacterial}%
  \BibitemOpen
  \bibfield  {author} {\bibinfo {author} {\bibfnamefont {E.}~\bibnamefont
  {Lauga}},\ }\bibfield  {title} {\bibinfo {title} {Bacterial hydrodynamics},\
  }\href@noop {} {\bibfield  {journal} {\bibinfo  {journal} {Annual Review of
  Fluid Mechanics}\ }\textbf {\bibinfo {volume} {48}},\ \bibinfo {pages} {105}
  (\bibinfo {year} {2016})}\BibitemShut {NoStop}%
\bibitem [{\citenamefont {Dreyfus}\ \emph {et~al.}(2005)\citenamefont
  {Dreyfus}, \citenamefont {Baudry}, \citenamefont {Roper}, \citenamefont
  {Fermigier}, \citenamefont {Stone},\ and\ \citenamefont
  {Bibette}}]{dreyfus2005microscopic}%
  \BibitemOpen
  \bibfield  {author} {\bibinfo {author} {\bibfnamefont {R.}~\bibnamefont
  {Dreyfus}}, \bibinfo {author} {\bibfnamefont {J.}~\bibnamefont {Baudry}},
  \bibinfo {author} {\bibfnamefont {M.~L.}\ \bibnamefont {Roper}}, \bibinfo
  {author} {\bibfnamefont {M.}~\bibnamefont {Fermigier}}, \bibinfo {author}
  {\bibfnamefont {H.~A.}\ \bibnamefont {Stone}},\ and\ \bibinfo {author}
  {\bibfnamefont {J.}~\bibnamefont {Bibette}},\ }\bibfield  {title} {\bibinfo
  {title} {Microscopic artificial swimmers},\ }\href@noop {} {\bibfield
  {journal} {\bibinfo  {journal} {Nature}\ }\textbf {\bibinfo {volume} {437}},\
  \bibinfo {pages} {862} (\bibinfo {year} {2005})}\BibitemShut {NoStop}%
\bibitem [{\citenamefont {Klotsa}\ \emph {et~al.}(2015)\citenamefont {Klotsa},
  \citenamefont {Baldwin}, \citenamefont {Hill},\ and\ \citenamefont
  {Swift}}]{klotsa2015propulsion}%
  \BibitemOpen
  \bibfield  {author} {\bibinfo {author} {\bibfnamefont {D.}~\bibnamefont
  {Klotsa}}, \bibinfo {author} {\bibfnamefont {K.~A.}\ \bibnamefont {Baldwin}},
  \bibinfo {author} {\bibfnamefont {R.~J.}\ \bibnamefont {Hill}},\ and\
  \bibinfo {author} {\bibfnamefont {M.~R.}\ \bibnamefont {Swift}},\ }\bibfield
  {title} {\bibinfo {title} {Propulsion of a two-sphere swimmer},\ }\href@noop
  {} {\bibfield  {journal} {\bibinfo  {journal} {Physical Review Letters}\
  }\textbf {\bibinfo {volume} {115}},\ \bibinfo {pages} {248102} (\bibinfo
  {year} {2015})}\BibitemShut {NoStop}%
\bibitem [{\citenamefont {Derr}\ \emph {et~al.}(2022)\citenamefont {Derr},
  \citenamefont {Dombrowski}, \citenamefont {Rycroft},\ and\ \citenamefont
  {Klotsa}}]{derr2022reciprocal}%
  \BibitemOpen
  \bibfield  {author} {\bibinfo {author} {\bibfnamefont {N.~J.}\ \bibnamefont
  {Derr}}, \bibinfo {author} {\bibfnamefont {T.}~\bibnamefont {Dombrowski}},
  \bibinfo {author} {\bibfnamefont {C.~H.}\ \bibnamefont {Rycroft}},\ and\
  \bibinfo {author} {\bibfnamefont {D.}~\bibnamefont {Klotsa}},\ }\bibfield
  {title} {\bibinfo {title} {Reciprocal swimming at intermediate reynolds
  number},\ }\href@noop {} {\bibfield  {journal} {\bibinfo  {journal} {Journal
  of Fluid Mechanics}\ }\textbf {\bibinfo {volume} {952}},\ \bibinfo {pages}
  {A8} (\bibinfo {year} {2022})}\BibitemShut {NoStop}%
\bibitem [{\citenamefont {Klotsa}(2019)}]{klotsa2019above}%
  \BibitemOpen
  \bibfield  {author} {\bibinfo {author} {\bibfnamefont {D.}~\bibnamefont
  {Klotsa}},\ }\bibfield  {title} {\bibinfo {title} {As above, so below, and
  also in between: mesoscale active matter in fluids},\ }\href@noop {}
  {\bibfield  {journal} {\bibinfo  {journal} {Soft Matter}\ }\textbf {\bibinfo
  {volume} {15}},\ \bibinfo {pages} {8946} (\bibinfo {year}
  {2019})}\BibitemShut {NoStop}%
\bibitem [{\citenamefont {L{\"o}wen}(2020)}]{lowen2020inertial}%
  \BibitemOpen
  \bibfield  {author} {\bibinfo {author} {\bibfnamefont {H.}~\bibnamefont
  {L{\"o}wen}},\ }\bibfield  {title} {\bibinfo {title} {Inertial effects of
  self-propelled particles: From active brownian to active langevin motion},\
  }\href@noop {} {\bibfield  {journal} {\bibinfo  {journal} {The Journal of
  Chemical Physics}\ }\textbf {\bibinfo {volume} {152}},\ \bibinfo {pages}
  {040901} (\bibinfo {year} {2020})}\BibitemShut {NoStop}%
\bibitem [{\citenamefont {Chen}\ \emph {et~al.}(2024)\citenamefont {Chen},
  \citenamefont {Weady}, \citenamefont {Atis}, \citenamefont {Matsuzawa},
  \citenamefont {Shelley},\ and\ \citenamefont {Irvine}}]{chen2024self}%
  \BibitemOpen
  \bibfield  {author} {\bibinfo {author} {\bibfnamefont {P.}~\bibnamefont
  {Chen}}, \bibinfo {author} {\bibfnamefont {S.}~\bibnamefont {Weady}},
  \bibinfo {author} {\bibfnamefont {S.}~\bibnamefont {Atis}}, \bibinfo {author}
  {\bibfnamefont {T.}~\bibnamefont {Matsuzawa}}, \bibinfo {author}
  {\bibfnamefont {M.~J.}\ \bibnamefont {Shelley}},\ and\ \bibinfo {author}
  {\bibfnamefont {W.~T.}\ \bibnamefont {Irvine}},\ }\bibfield  {title}
  {\bibinfo {title} {Self-propulsion, flocking and chiral active phases from
  particles spinning at intermediate reynolds numbers},\ }\href@noop {}
  {\bibfield  {journal} {\bibinfo  {journal} {Nature Physics}\ ,\ \bibinfo
  {pages} {1}} (\bibinfo {year} {2024})}\BibitemShut {NoStop}%
\bibitem [{\citenamefont {Klumpp}\ \emph {et~al.}(2019)\citenamefont {Klumpp},
  \citenamefont {Lef{\`e}vre}, \citenamefont {Bennet},\ and\ \citenamefont
  {Faivre}}]{klumpp2019swimming}%
  \BibitemOpen
  \bibfield  {author} {\bibinfo {author} {\bibfnamefont {S.}~\bibnamefont
  {Klumpp}}, \bibinfo {author} {\bibfnamefont {C.~T.}\ \bibnamefont
  {Lef{\`e}vre}}, \bibinfo {author} {\bibfnamefont {M.}~\bibnamefont
  {Bennet}},\ and\ \bibinfo {author} {\bibfnamefont {D.}~\bibnamefont
  {Faivre}},\ }\bibfield  {title} {\bibinfo {title} {Swimming with magnets:
  from biological organisms to synthetic devices},\ }\href@noop {} {\bibfield
  {journal} {\bibinfo  {journal} {Physics Reports}\ }\textbf {\bibinfo {volume}
  {789}},\ \bibinfo {pages} {1} (\bibinfo {year} {2019})}\BibitemShut {NoStop}%
\bibitem [{\citenamefont {Purcell}(1997)}]{purcell1997efficiency}%
  \BibitemOpen
  \bibfield  {author} {\bibinfo {author} {\bibfnamefont {E.~M.}\ \bibnamefont
  {Purcell}},\ }\bibfield  {title} {\bibinfo {title} {The efficiency of
  propulsion by a rotating flagellum},\ }\href@noop {} {\bibfield  {journal}
  {\bibinfo  {journal} {Proceedings of the National Academy of Sciences}\
  }\textbf {\bibinfo {volume} {94}},\ \bibinfo {pages} {11307} (\bibinfo {year}
  {1997})}\BibitemShut {NoStop}%
\bibitem [{\citenamefont {Rodenborn}\ \emph {et~al.}(2013)\citenamefont
  {Rodenborn}, \citenamefont {Chen}, \citenamefont {Swinney}, \citenamefont
  {Liu},\ and\ \citenamefont {Zhang}}]{rodenborn2013propulsion}%
  \BibitemOpen
  \bibfield  {author} {\bibinfo {author} {\bibfnamefont {B.}~\bibnamefont
  {Rodenborn}}, \bibinfo {author} {\bibfnamefont {C.-H.}\ \bibnamefont {Chen}},
  \bibinfo {author} {\bibfnamefont {H.~L.}\ \bibnamefont {Swinney}}, \bibinfo
  {author} {\bibfnamefont {B.}~\bibnamefont {Liu}},\ and\ \bibinfo {author}
  {\bibfnamefont {H.}~\bibnamefont {Zhang}},\ }\bibfield  {title} {\bibinfo
  {title} {Propulsion of microorganisms by a helical flagellum},\ }\href@noop
  {} {\bibfield  {journal} {\bibinfo  {journal} {Proceedings of the National
  Academy of Sciences}\ }\textbf {\bibinfo {volume} {110}},\ \bibinfo {pages}
  {E338} (\bibinfo {year} {2013})}\BibitemShut {NoStop}%
\bibitem [{\citenamefont {Bricard}\ \emph {et~al.}(2013)\citenamefont
  {Bricard}, \citenamefont {Caussin}, \citenamefont {Desreumaux}, \citenamefont
  {Dauchot},\ and\ \citenamefont {Bartolo}}]{bricard2013emergence}%
  \BibitemOpen
  \bibfield  {author} {\bibinfo {author} {\bibfnamefont {A.}~\bibnamefont
  {Bricard}}, \bibinfo {author} {\bibfnamefont {J.-B.}\ \bibnamefont
  {Caussin}}, \bibinfo {author} {\bibfnamefont {N.}~\bibnamefont {Desreumaux}},
  \bibinfo {author} {\bibfnamefont {O.}~\bibnamefont {Dauchot}},\ and\ \bibinfo
  {author} {\bibfnamefont {D.}~\bibnamefont {Bartolo}},\ }\bibfield  {title}
  {\bibinfo {title} {Emergence of macroscopic directed motion in populations of
  motile colloids},\ }\href@noop {} {\bibfield  {journal} {\bibinfo  {journal}
  {Nature}\ }\textbf {\bibinfo {volume} {503}},\ \bibinfo {pages} {95}
  (\bibinfo {year} {2013})}\BibitemShut {NoStop}%
\bibitem [{\citenamefont {Kaiser}\ \emph {et~al.}(2017)\citenamefont {Kaiser},
  \citenamefont {Snezhko},\ and\ \citenamefont {Aranson}}]{kaiser2017flocking}%
  \BibitemOpen
  \bibfield  {author} {\bibinfo {author} {\bibfnamefont {A.}~\bibnamefont
  {Kaiser}}, \bibinfo {author} {\bibfnamefont {A.}~\bibnamefont {Snezhko}},\
  and\ \bibinfo {author} {\bibfnamefont {I.~S.}\ \bibnamefont {Aranson}},\
  }\bibfield  {title} {\bibinfo {title} {Flocking ferromagnetic colloids},\
  }\href@noop {} {\bibfield  {journal} {\bibinfo  {journal} {Science Advances}\
  }\textbf {\bibinfo {volume} {3}},\ \bibinfo {pages} {e1601469} (\bibinfo
  {year} {2017})}\BibitemShut {NoStop}%
\bibitem [{\citenamefont {Driscoll}\ \emph {et~al.}(2017)\citenamefont
  {Driscoll}, \citenamefont {Delmotte}, \citenamefont {Youssef}, \citenamefont
  {Sacanna}, \citenamefont {Donev},\ and\ \citenamefont
  {Chaikin}}]{driscoll2017unstable}%
  \BibitemOpen
  \bibfield  {author} {\bibinfo {author} {\bibfnamefont {M.}~\bibnamefont
  {Driscoll}}, \bibinfo {author} {\bibfnamefont {B.}~\bibnamefont {Delmotte}},
  \bibinfo {author} {\bibfnamefont {M.}~\bibnamefont {Youssef}}, \bibinfo
  {author} {\bibfnamefont {S.}~\bibnamefont {Sacanna}}, \bibinfo {author}
  {\bibfnamefont {A.}~\bibnamefont {Donev}},\ and\ \bibinfo {author}
  {\bibfnamefont {P.}~\bibnamefont {Chaikin}},\ }\bibfield  {title} {\bibinfo
  {title} {Unstable fronts and motile structures formed by microrollers},\
  }\href@noop {} {\bibfield  {journal} {\bibinfo  {journal} {Nature Physics}\
  }\textbf {\bibinfo {volume} {13}},\ \bibinfo {pages} {375} (\bibinfo {year}
  {2017})}\BibitemShut {NoStop}%
\bibitem [{\citenamefont {Alapan}\ \emph {et~al.}(2020)\citenamefont {Alapan},
  \citenamefont {Bozuyuk}, \citenamefont {Erkoc}, \citenamefont {Karacakol},\
  and\ \citenamefont {Sitti}}]{alapan2020multifunctional}%
  \BibitemOpen
  \bibfield  {author} {\bibinfo {author} {\bibfnamefont {Y.}~\bibnamefont
  {Alapan}}, \bibinfo {author} {\bibfnamefont {U.}~\bibnamefont {Bozuyuk}},
  \bibinfo {author} {\bibfnamefont {P.}~\bibnamefont {Erkoc}}, \bibinfo
  {author} {\bibfnamefont {A.~C.}\ \bibnamefont {Karacakol}},\ and\ \bibinfo
  {author} {\bibfnamefont {M.}~\bibnamefont {Sitti}},\ }\bibfield  {title}
  {\bibinfo {title} {Multifunctional surface microrollers for targeted cargo
  delivery in physiological blood flow},\ }\href@noop {} {\bibfield  {journal}
  {\bibinfo  {journal} {Science Robotics}\ }\textbf {\bibinfo {volume} {5}},\
  \bibinfo {pages} {eaba5726} (\bibinfo {year} {2020})}\BibitemShut {NoStop}%
\bibitem [{\citenamefont {Grzybowski}\ \emph {et~al.}(2000)\citenamefont
  {Grzybowski}, \citenamefont {Stone},\ and\ \citenamefont
  {Whitesides}}]{grzybowski2000dynamic}%
  \BibitemOpen
  \bibfield  {author} {\bibinfo {author} {\bibfnamefont {B.~A.}\ \bibnamefont
  {Grzybowski}}, \bibinfo {author} {\bibfnamefont {H.~A.}\ \bibnamefont
  {Stone}},\ and\ \bibinfo {author} {\bibfnamefont {G.~M.}\ \bibnamefont
  {Whitesides}},\ }\bibfield  {title} {\bibinfo {title} {Dynamic self-assembly
  of magnetized, millimetre-sized objects rotating at a liquid--air
  interface},\ }\href@noop {} {\bibfield  {journal} {\bibinfo  {journal}
  {Nature}\ }\textbf {\bibinfo {volume} {405}},\ \bibinfo {pages} {1033}
  (\bibinfo {year} {2000})}\BibitemShut {NoStop}%
\bibitem [{\citenamefont {Goto}\ and\ \citenamefont
  {Tanaka}(2015)}]{goto2015purely}%
  \BibitemOpen
  \bibfield  {author} {\bibinfo {author} {\bibfnamefont {Y.}~\bibnamefont
  {Goto}}\ and\ \bibinfo {author} {\bibfnamefont {H.}~\bibnamefont {Tanaka}},\
  }\bibfield  {title} {\bibinfo {title} {Purely hydrodynamic ordering of
  rotating disks at a finite reynolds number},\ }\href@noop {} {\bibfield
  {journal} {\bibinfo  {journal} {Nature Communications}\ }\textbf {\bibinfo
  {volume} {6}},\ \bibinfo {pages} {5994} (\bibinfo {year} {2015})}\BibitemShut
  {NoStop}%
\bibitem [{\citenamefont {Fang}\ \emph {et~al.}(2020)\citenamefont {Fang},
  \citenamefont {Ham}, \citenamefont {Qiao},\ and\ \citenamefont
  {Tao}}]{fang2020magnetic}%
  \BibitemOpen
  \bibfield  {author} {\bibinfo {author} {\bibfnamefont {W.-Z.}\ \bibnamefont
  {Fang}}, \bibinfo {author} {\bibfnamefont {S.}~\bibnamefont {Ham}}, \bibinfo
  {author} {\bibfnamefont {R.}~\bibnamefont {Qiao}},\ and\ \bibinfo {author}
  {\bibfnamefont {W.-Q.}\ \bibnamefont {Tao}},\ }\bibfield  {title} {\bibinfo
  {title} {Magnetic actuation of surface walkers: The effects of confinement
  and inertia},\ }\href@noop {} {\bibfield  {journal} {\bibinfo  {journal}
  {Langmuir}\ }\textbf {\bibinfo {volume} {36}},\ \bibinfo {pages} {7046}
  (\bibinfo {year} {2020})}\BibitemShut {NoStop}%
\bibitem [{\citenamefont {Shen}\ and\ \citenamefont
  {Lintuvuori}(2023)}]{shen2023collective}%
  \BibitemOpen
  \bibfield  {author} {\bibinfo {author} {\bibfnamefont {Z.}~\bibnamefont
  {Shen}}\ and\ \bibinfo {author} {\bibfnamefont {J.~S.}\ \bibnamefont
  {Lintuvuori}},\ }\bibfield  {title} {\bibinfo {title} {Collective flows drive
  cavitation in spinner monolayers},\ }\href@noop {} {\bibfield  {journal}
  {\bibinfo  {journal} {Physical Review Letters}\ }\textbf {\bibinfo {volume}
  {130}},\ \bibinfo {pages} {188202} (\bibinfo {year} {2023})}\BibitemShut
  {NoStop}%
\bibitem [{\citenamefont {Nadal}\ \emph {et~al.}(2014)\citenamefont {Nadal},
  \citenamefont {Pak}, \citenamefont {Zhu}, \citenamefont {Brandt},\ and\
  \citenamefont {Lauga}}]{nadal2014rotational}%
  \BibitemOpen
  \bibfield  {author} {\bibinfo {author} {\bibfnamefont {F.}~\bibnamefont
  {Nadal}}, \bibinfo {author} {\bibfnamefont {O.~S.}\ \bibnamefont {Pak}},
  \bibinfo {author} {\bibfnamefont {L.}~\bibnamefont {Zhu}}, \bibinfo {author}
  {\bibfnamefont {L.}~\bibnamefont {Brandt}},\ and\ \bibinfo {author}
  {\bibfnamefont {E.}~\bibnamefont {Lauga}},\ }\bibfield  {title} {\bibinfo
  {title} {Rotational propulsion enabled by inertia},\ }\href@noop {}
  {\bibfield  {journal} {\bibinfo  {journal} {The European Physical Journal E}\
  }\textbf {\bibinfo {volume} {37}},\ \bibinfo {pages} {1} (\bibinfo {year}
  {2014})}\BibitemShut {NoStop}%
\bibitem [{\citenamefont {Bickley}(1938)}]{bickley1938lxv}%
  \BibitemOpen
  \bibfield  {author} {\bibinfo {author} {\bibfnamefont {W.}~\bibnamefont
  {Bickley}},\ }\bibfield  {title} {\bibinfo {title} {Lxv. the secondary flow
  due to a sphere rotating in a viscous fluid},\ }\href@noop {} {\bibfield
  {journal} {\bibinfo  {journal} {The London, Edinburgh, and Dublin
  Philosophical Magazine and Journal of Science}\ }\textbf {\bibinfo {volume}
  {25}},\ \bibinfo {pages} {746} (\bibinfo {year} {1938})}\BibitemShut
  {NoStop}%
\bibitem [{\citenamefont {Liu}\ and\ \citenamefont
  {Prosperetti}(2010)}]{liu2010wall}%
  \BibitemOpen
  \bibfield  {author} {\bibinfo {author} {\bibfnamefont {Q.}~\bibnamefont
  {Liu}}\ and\ \bibinfo {author} {\bibfnamefont {A.}~\bibnamefont
  {Prosperetti}},\ }\bibfield  {title} {\bibinfo {title} {Wall effects on a
  rotating sphere},\ }\href@noop {} {\bibfield  {journal} {\bibinfo  {journal}
  {Journal of Fluid Mechanics}\ }\textbf {\bibinfo {volume} {657}},\ \bibinfo
  {pages} {1} (\bibinfo {year} {2010})}\BibitemShut {NoStop}%
\bibitem [{\citenamefont {kevinstratford}\ \emph {et~al.}(2024)\citenamefont
  {kevinstratford}, \citenamefont {Henrich}, \citenamefont {jlintuvuori},
  \citenamefont {dmarendu}, \citenamefont {qikaifzj}, \citenamefont
  {austin1997}, \citenamefont {shanCHEN123}, \citenamefont {ludwig cf},
  \citenamefont {jurijsab}, \citenamefont {Leyva},\ and\ \citenamefont
  {sumeshpt}}]{kevinstratford_2024_12822477}%
  \BibitemOpen
  \bibfield  {author} {\bibinfo {author} {\bibnamefont {kevinstratford}},
  \bibinfo {author} {\bibfnamefont {O.}~\bibnamefont {Henrich}}, \bibinfo
  {author} {\bibnamefont {jlintuvuori}}, \bibinfo {author} {\bibnamefont
  {dmarendu}}, \bibinfo {author} {\bibnamefont {qikaifzj}}, \bibinfo {author}
  {\bibnamefont {austin1997}}, \bibinfo {author} {\bibnamefont {shanCHEN123}},
  \bibinfo {author} {\bibnamefont {ludwig cf}}, \bibinfo {author} {\bibnamefont
  {jurijsab}}, \bibinfo {author} {\bibfnamefont {S.~G.}\ \bibnamefont
  {Leyva}},\ and\ \bibinfo {author} {\bibnamefont {sumeshpt}},\ }\href
  {https://doi.org/10.5281/zenodo.12822477} {\bibinfo {title}
  {ludwig-cf/ludwig: Ludwig 0.22.0}} (\bibinfo {year} {2024})\BibitemShut
  {NoStop}%
\bibitem [{\citenamefont {Shen}\ \emph {et~al.}(2018)\citenamefont {Shen},
  \citenamefont {W{\"u}rger},\ and\ \citenamefont
  {Lintuvuori}}]{shen2018hydrodynamic}%
  \BibitemOpen
  \bibfield  {author} {\bibinfo {author} {\bibfnamefont {Z.}~\bibnamefont
  {Shen}}, \bibinfo {author} {\bibfnamefont {A.}~\bibnamefont {W{\"u}rger}},\
  and\ \bibinfo {author} {\bibfnamefont {J.~S.}\ \bibnamefont {Lintuvuori}},\
  }\bibfield  {title} {\bibinfo {title} {Hydrodynamic interaction of a
  self-propelling particle with a wall: Comparison between an active janus
  particle and a squirmer model},\ }\href@noop {} {\bibfield  {journal}
  {\bibinfo  {journal} {The European Physical Journal E}\ }\textbf {\bibinfo
  {volume} {41}},\ \bibinfo {pages} {1} (\bibinfo {year} {2018})}\BibitemShut
  {NoStop}%
\bibitem [{\citenamefont {Shen}\ \emph {et~al.}(2019)\citenamefont {Shen},
  \citenamefont {W{\"u}rger},\ and\ \citenamefont
  {Lintuvuori}}]{shen2019hydrodynamic}%
  \BibitemOpen
  \bibfield  {author} {\bibinfo {author} {\bibfnamefont {Z.}~\bibnamefont
  {Shen}}, \bibinfo {author} {\bibfnamefont {A.}~\bibnamefont {W{\"u}rger}},\
  and\ \bibinfo {author} {\bibfnamefont {J.~S.}\ \bibnamefont {Lintuvuori}},\
  }\bibfield  {title} {\bibinfo {title} {Hydrodynamic self-assembly of active
  colloids: chiral spinners and dynamic crystals},\ }\href@noop {} {\bibfield
  {journal} {\bibinfo  {journal} {Soft Matter}\ }\textbf {\bibinfo {volume}
  {15}},\ \bibinfo {pages} {1508} (\bibinfo {year} {2019})}\BibitemShut
  {NoStop}%
\bibitem [{\citenamefont {Climent}\ \emph {et~al.}(2006)\citenamefont
  {Climent}, \citenamefont {Yeo}, \citenamefont {Maxey},\ and\ \citenamefont
  {Karniadakis}}]{climent2007dynamic}%
  \BibitemOpen
  \bibfield  {author} {\bibinfo {author} {\bibfnamefont {E.}~\bibnamefont
  {Climent}}, \bibinfo {author} {\bibfnamefont {K.}~\bibnamefont {Yeo}},
  \bibinfo {author} {\bibfnamefont {M.~R.}\ \bibnamefont {Maxey}},\ and\
  \bibinfo {author} {\bibfnamefont {G.~E.}\ \bibnamefont {Karniadakis}},\
  }\bibfield  {title} {\bibinfo {title} {Dynamic self-assembly of spinning
  particles},\ }\href {https://doi.org/10.1115/1.2436587} {\bibfield  {journal}
  {\bibinfo  {journal} {Journal of Fluids Engineering}\ }\textbf {\bibinfo
  {volume} {129}},\ \bibinfo {pages} {379} (\bibinfo {year}
  {2006})}\BibitemShut {NoStop}%
\bibitem [{\citenamefont {Shen}\ and\ \citenamefont
  {Lintuvuori}(2020)}]{shen2020hydrodynamic}%
  \BibitemOpen
  \bibfield  {author} {\bibinfo {author} {\bibfnamefont {Z.}~\bibnamefont
  {Shen}}\ and\ \bibinfo {author} {\bibfnamefont {J.~S.}\ \bibnamefont
  {Lintuvuori}},\ }\bibfield  {title} {\bibinfo {title} {Hydrodynamic
  clustering and emergent phase separation of spherical spinners},\ }\href@noop
  {} {\bibfield  {journal} {\bibinfo  {journal} {Physical Review Research}\
  }\textbf {\bibinfo {volume} {2}},\ \bibinfo {pages} {013358} (\bibinfo {year}
  {2020})}\BibitemShut {NoStop}%
\bibitem [{\citenamefont {Kroo}\ \emph {et~al.}(2022)\citenamefont {Kroo},
  \citenamefont {Binagia}, \citenamefont {Eckman}, \citenamefont {Prakash},\
  and\ \citenamefont {Shaqfeh}}]{kroo2022freely}%
  \BibitemOpen
  \bibfield  {author} {\bibinfo {author} {\bibfnamefont {L.}~\bibnamefont
  {Kroo}}, \bibinfo {author} {\bibfnamefont {J.~P.}\ \bibnamefont {Binagia}},
  \bibinfo {author} {\bibfnamefont {N.}~\bibnamefont {Eckman}}, \bibinfo
  {author} {\bibfnamefont {M.}~\bibnamefont {Prakash}},\ and\ \bibinfo {author}
  {\bibfnamefont {E.~S.}\ \bibnamefont {Shaqfeh}},\ }\bibfield  {title}
  {\bibinfo {title} {A freely suspended robotic swimmer propelled by
  viscoelastic normal stresses},\ }\href@noop {} {\bibfield  {journal}
  {\bibinfo  {journal} {Journal of Fluid Mechanics}\ }\textbf {\bibinfo
  {volume} {944}},\ \bibinfo {pages} {A20} (\bibinfo {year}
  {2022})}\BibitemShut {NoStop}%
\bibitem [{\citenamefont {Binagia}\ and\ \citenamefont
  {Shaqfeh}(2021)}]{binagia2021self}%
  \BibitemOpen
  \bibfield  {author} {\bibinfo {author} {\bibfnamefont {J.~P.}\ \bibnamefont
  {Binagia}}\ and\ \bibinfo {author} {\bibfnamefont {E.~S.}\ \bibnamefont
  {Shaqfeh}},\ }\bibfield  {title} {\bibinfo {title} {Self-propulsion of a
  freely suspended swimmer by a swirling tail in a viscoelastic fluid},\
  }\href@noop {} {\bibfield  {journal} {\bibinfo  {journal} {Physical Review
  Fluids}\ }\textbf {\bibinfo {volume} {6}},\ \bibinfo {pages} {053301}
  (\bibinfo {year} {2021})}\BibitemShut {NoStop}%
\bibitem [{\citenamefont {Goychuk}\ \emph {et~al.}(2014)\citenamefont
  {Goychuk}, \citenamefont {Kharchenko},\ and\ \citenamefont
  {Metzler}}]{goychuk2014molecular}%
  \BibitemOpen
  \bibfield  {author} {\bibinfo {author} {\bibfnamefont {I.}~\bibnamefont
  {Goychuk}}, \bibinfo {author} {\bibfnamefont {V.~O.}\ \bibnamefont
  {Kharchenko}},\ and\ \bibinfo {author} {\bibfnamefont {R.}~\bibnamefont
  {Metzler}},\ }\bibfield  {title} {\bibinfo {title} {Molecular motors pulling
  cargos in the viscoelastic cytosol: how power strokes beat subdiffusion},\
  }\href@noop {} {\bibfield  {journal} {\bibinfo  {journal} {Physical Chemistry
  Chemical Physics}\ }\textbf {\bibinfo {volume} {16}},\ \bibinfo {pages}
  {16524} (\bibinfo {year} {2014})}\BibitemShut {NoStop}%
\bibitem [{\citenamefont {Li}\ \emph {et~al.}(2021)\citenamefont {Li},
  \citenamefont {Lauga},\ and\ \citenamefont {Ardekani}}]{li2021microswimming}%
  \BibitemOpen
  \bibfield  {author} {\bibinfo {author} {\bibfnamefont {G.}~\bibnamefont
  {Li}}, \bibinfo {author} {\bibfnamefont {E.}~\bibnamefont {Lauga}},\ and\
  \bibinfo {author} {\bibfnamefont {A.~M.}\ \bibnamefont {Ardekani}},\
  }\bibfield  {title} {\bibinfo {title} {Microswimming in viscoelastic
  fluids},\ }\href@noop {} {\bibfield  {journal} {\bibinfo  {journal} {Journal
  of Non-Newtonian Fluid Mechanics}\ }\textbf {\bibinfo {volume} {297}},\
  \bibinfo {pages} {104655} (\bibinfo {year} {2021})}\BibitemShut {NoStop}%
\bibitem [{\citenamefont {Theeyancheri}\ \emph {et~al.}(2022)\citenamefont
  {Theeyancheri}, \citenamefont {Sahoo}, \citenamefont {Kumar},\ and\
  \citenamefont {Chakrabarti}}]{theeyancheri2022silico}%
  \BibitemOpen
  \bibfield  {author} {\bibinfo {author} {\bibfnamefont {L.}~\bibnamefont
  {Theeyancheri}}, \bibinfo {author} {\bibfnamefont {R.}~\bibnamefont {Sahoo}},
  \bibinfo {author} {\bibfnamefont {P.}~\bibnamefont {Kumar}},\ and\ \bibinfo
  {author} {\bibfnamefont {R.}~\bibnamefont {Chakrabarti}},\ }\bibfield
  {title} {\bibinfo {title} {In silico studies of active probe dynamics in
  crowded media},\ }\href@noop {} {\bibfield  {journal} {\bibinfo  {journal}
  {ACS omega}\ }\textbf {\bibinfo {volume} {7}},\ \bibinfo {pages} {33637}
  (\bibinfo {year} {2022})}\BibitemShut {NoStop}%
\end{thebibliography}%

\end{document}